\documentclass[conference]{IEEEtran}

\usepackage{cite}
\usepackage{amsmath,amssymb,amsfonts}
\usepackage{algorithmic}
\usepackage{graphicx}
\usepackage{textcomp}
\usepackage{xcolor}

\def\BibTeX{{\rm B\kern-.05em{\sc i\kern-.025em b}\kern-.08em
    T\kern-.1667em\lower.7ex\hbox{E}\kern-.125emX}}
\begin{document}

\title{AMT-APC: Automatic Piano Cover by Fine-Tuning an Automatic Music Transcription Model}

\author{
\IEEEauthorblockN{Kazuma Komiya}
\IEEEauthorblockA{
  \textit{Faculty of Data Science} \\ 
  \textit{Musashino University} \\
  Tokyo, Japan \\
  ORCID: 0009-0004-2477-0162
}
\and
\IEEEauthorblockN{Yoshihisa Fukuhara}
\IEEEauthorblockA{
  \textit{Faculty of Data Science} \\ 
  \textit{Musashino University} \\
  Tokyo, Japan \\
  ORCID: 0000-0003-3458-7463
}
}

\maketitle

\begin{abstract}
There have been several studies on automatically generating piano covers, and recent advancements in deep learning have enabled the creation of more sophisticated covers. However, existing automatic piano cover models still have room for improvement in terms of expressiveness and fidelity to the original. To address these issues, we propose a learning algorithm called AMT-APC, which leverages the capabilities of automatic music transcription models. By utilizing the strengths of well-established automatic music transcription models, we aim to improve the accuracy of piano cover generation. Our experiments demonstrate that the AMT-APC model reproduces original tracks more accurately than any existing models.

\end{abstract}


\section{Introduction}\label{sec:intro}
The piano is one of the most popular instruments in the world, and people enjoy it in various ways, such as through performance, listening, and creation. When performing a favorite song on the piano, the song must be covered for piano performance. However, to create a piano cover, one must first accurately listen to the sounds from the original track, understand the musical structure, such as the melody and chord progression, and then transcribe them into a playable sheet music format. Unfortunately, executing this process requires a high level of musical knowledge and technical skill, making it difficult for everyone to accomplish easily.

In recent years, the development of deep learning has had a significant impact across various fields, such as language modeling and computer vision \cite{transformer,GPT1,BERT,GAN,Rombach2022}. The music domain is no exception, with deep learning contributing to various areas, including music generation \cite{jukebox,music-transformer,REMI,multitrack-music-transformer} and automatic music transcription (AMT) \cite{onsets-frames,MT3,hawthorne2021,hft-transformer}. Approaches using deep learning have also been proposed for automatic piano covers (APC) \cite{pop2piano,picogen1,picogen2}, but these models still face limitations in terms of fidelity to the original track and expressiveness.

To address these challenges, we propose an algorithm called AMT-APC, which trains an APC model by fine-tuning an AMT model. This algorithm consists of pre-training on the AMT task and fine-tuning for piano cover generation. Our experiments demonstrate that the APC model trained with this algorithm can reproduce original songs with high accuracy. The source code and demo audio used in this study are shared on the project page\footnote{https://310hz.github.io/amt-apc/}.

\section{Background}\label{sec:bg}
\subsection{Automatic Music Transcription}\label{subsec:AMT}
Automatic music transcription is a task that involves predicting the notes being played from a performance audio recording, and it is abbreviated as AMT. In the onsets-frames model \cite{onsets-frames}, two types of data, onsets and frames, are represented as a two-dimensional matrix similar to a piano roll, enabling the accurate prediction of piano notes from the given audio input. By predicting the characteristic moments of note onsets through a separate layer, this model allows for more precise note prediction. There are many piano AMT models that adopt the idea of onsets-frames \cite{hft-transformer,kim2019,kong2021}.

\subsection{Automatic Piano Cover}\label{subsec:APC}
Automatic piano cover is the task of generating a piano cover in MIDI format from an original audio track. In this paper, we refer to this task as APC.

In pop2piano \cite{pop2piano}, an APC model is obtained by training T5 \cite{T5} with paired data consisting of original audio tracks and piano-covered MIDI. However, since pop2piano quantizes all notes into eighth notes, it cannot reproduce more complex rhythms, such as those with finer notes or triplets.

PiCoGen1 \cite{picogen1} implemented APC training without using paired data by employing a lead sheet as an intermediary. However, lead sheets generally contain only essential information representing the musical structure, such as melody and chord progressions. Since a piano cover typically incorporates not only these key elements but also other aspects like drum patterns, timbre, dynamics, and grace notes, PiCoGen1 fails to sufficiently account for all the elements necessary for a piano cover.

PiCoGen2 \cite{picogen2}, similar to pop2piano, achieved APC training using paired data. It leveraged SheetSage \cite{sheetsage}, which was pre-trained for transcription tasks, to extract intermediate representations from the original audio. However, since the parameters of SheetSage are fixed during the training of piano covers, it may not be able to learn all the essential elements required for creating a piano cover, as mentioned earlier.

\section{Methodology}\label{sec:method}
\begin{figure}[tb]
\centerline{\includegraphics[width=\linewidth]{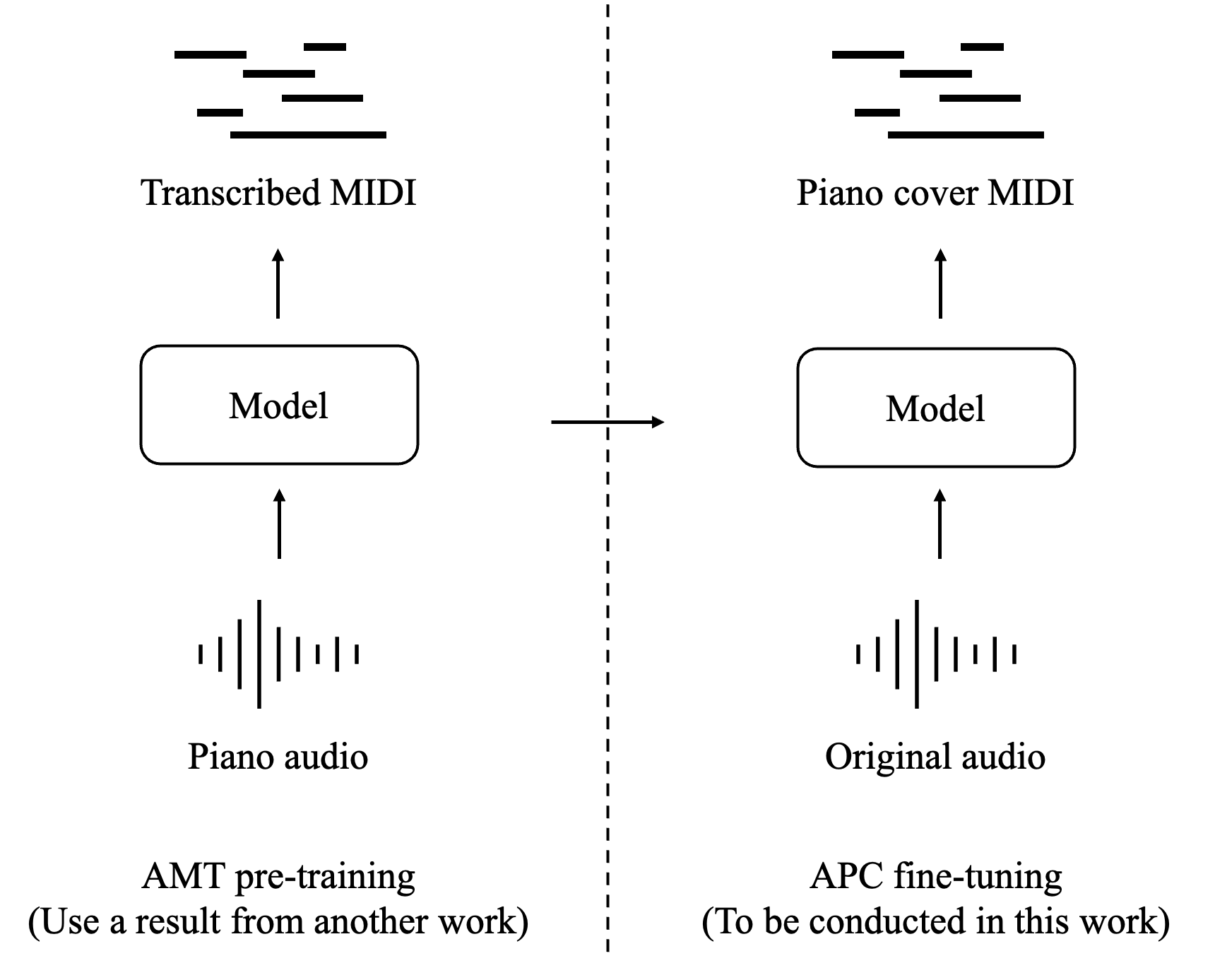}}
\caption{Overview of AMT-APC. It consists of AMT pre-training and APC fine-tuning. A pre-trained AMT model is used, and fine-tuning is performed in this study.}
\label{fig:overview}
\end{figure}

We propose a learning algorithm called AMT-APC as a solution to the challenges faced by existing APC models. Focusing on the AMT model's ability to "accurately capture sounds," we leverage this capability in the APC task to develop a model that can precisely reproduce the acoustic characteristics of the original track. Figure \ref{fig:overview} provides an overview of the proposed method. In AMT-APC, the same model architecture and MIDI representation as the previously studied AMT models are used for the APC model, and the parameters of a pre-trained AMT model are employed as the initial values for the APC model's parameters. This algorithm can be interpreted as consisting of pre-training on the AMT task followed by fine-tuning for piano cover generation.

\subsection{Base AMT Model}\label{subsec:PT}
It is essential to select a piano AMT model that has been thoroughly researched and proven to perform well as the foundation for the APC model. We adopt hFT-Transformer \cite{hft-transformer} as the base AMT model, as it has demonstrated high performance in piano AMT tasks.

Pop2Piano uses MT3 \cite{MT3} as its base and utilizes a one-dimensional MIDI-like representation. In contrast, we adopt hFT-Transformer, which employs a two-dimensional piano-roll-like MIDI representation. This choice is made not only because hFT-Transformer performs better in piano AMT tasks but also to incorporate an "approach that focuses on where the sounds are played," enabled by the two-dimensional MIDI representation (explained in section \ref{subsec:ft}).

The hFT-Transformer is divided into two main hierarchies, and after receiving the mel-spectrogram of the input audio, it outputs four matrices—onsets, offsets, frames, and velocities—within each hierarchy. In other words, it outputs a total of eight matrices. In this work, to avoid unnecessary complexity, we do not train the offsets and ignore it during inference. During training, the loss is measured based on the outputs from both layers, while for inference, the output from the second layer is used. The onsets and frames are represented as 512x88 matrices, with each element taking a value between 0 and 1. The velocities are represented as 512x88x128 data, where each element contains a 128-dimensional vector representing the probability distribution of velocity values. The representation of onsets follows \cite{kong2021}, indicating the distance to the nearest onset.

The publicly available pre-trained hFT-Transformer model processes sequences of 128 frames (approximately 2 seconds) at a time, dividing the entire piece into segments and processing each independently. For APC, we determined that this length was insufficient and increased the sequence length to 512 frames. To train the additional positional embeddings introduced by this change, we performed additional fine-tuning on the MAESTRO dataset \cite{MAESTRO} for less than one epoch.

\subsection{Style Vector}\label{subsec:sv}
Even when covering a single piece of music, various covers can be imagined depending on the style. If the model is not provided with information about the cover style, it may become confused and unsure of which style to reflect in the MIDI output.

To address this, we propose a continuous vector representing the cover style, referred to as a style vector. While pop2piano assigned an ID to each arranger, we introduce continuous vectors to allow the model to learn more diverse expressions. By extracting a style vector from each piano cover and feeding it to the model alongside the original audio, the model can learn piano covers that take into account different styles. This is expected to enable the model to learn various styles while maintaining consistency across the entire piece.

\begin{figure}[tb]
\centerline{\includegraphics[width=\linewidth]{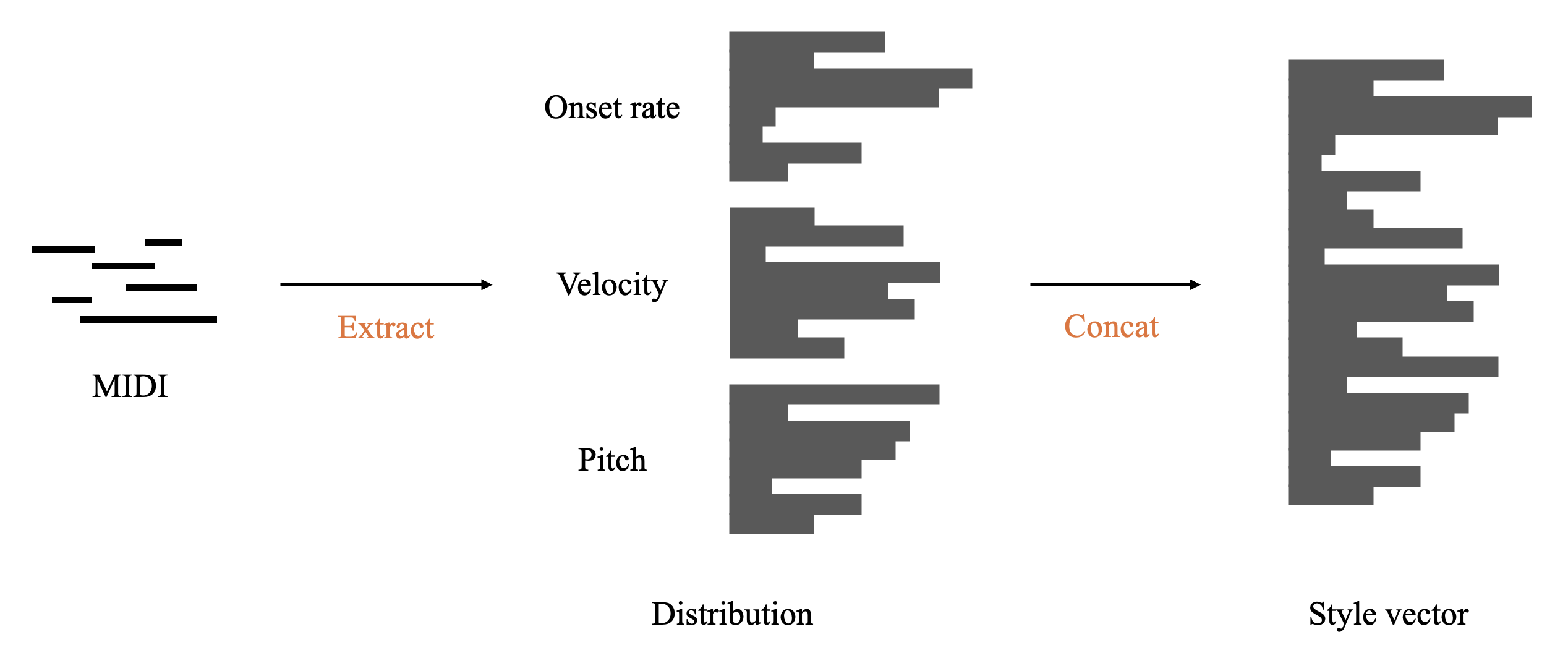}}
\caption{Method of extracting the style vector. Probability distributions related to onset rate, velocity, and pitch are extracted and combined to form a single vector.}
\label{fig:exsv}
\end{figure}

The method for extracting the style vector from MIDI is shown in Figure \ref{fig:exsv}. By extracting the following three probability distributions, which represent style-related features, and combining them, we obtain the style vector.

\begin{itemize}
  \item \textbf{Onset rate distribution}: The distribution of onset rates. Onset rate refers to the average number of onsets per frame within a segment. This concept is inspired by MuseMorphose \cite{musemorphose}.
  \item \textbf{Velocity distribution}: The distribution of velocities across the entire cover (128 levels).
  \item \textbf{Pitch distribution}: The distribution of pitches across the entire cover (88 levels).
\end{itemize}

These three features are extracted. When feeding this to the model, each distribution is first normalized to have a mean of 0 and a standard deviation of 1, then quantized into 8 levels using 7 bins.

\begin{align*}
  [-2,\, -4/3,\, -2/3,\, 0,\, 2/3,\, 4/3,\, 2]
\end{align*}

The three 8-dimensional vectors obtained are concatenated to form a 24-dimensional vector, which serves as the style vector for the cover.

Lastly, we explain how the style vectors are provided to the model. The hFT-Transformer we adopted is divided into two main hierarchies, with the first hierarchy consisting of an encoder and a decoder. The style vectors are added to all hidden states output by the encoder in this first hierarchy. Specifically, the 24-dimensional vector is mapped to the hidden state space $\mathbb{R}^Z$ via a fully-connected layer. After that, it is added to the hidden state $\boldsymbol{h}_{t,f}$ at time $t$ and frequency $f$ through a gate $\boldsymbol{r}_{t,f}$. This process is formalized as follows:

\begin{align}\label{eq:svlayer}
  \boldsymbol h_{\text{sv}} &= W\boldsymbol v_s + \boldsymbol b \\
  \boldsymbol r_{t, f} &= \text{gate}(\boldsymbol h_{t, f}) \\
  \boldsymbol {\tilde h}_{t, f} &= \boldsymbol r_{t, f}\boldsymbol h_{t, f} + (1 - \boldsymbol r_{t, f})\boldsymbol h_{\text{sv}}
\end{align}

where $\boldsymbol{h}_{\text{sv}} \in \mathbb{R}^Z$ is the linear map of the style vector $v_s \in \mathbb{R}^{24}$, and $\boldsymbol{h}_{t,f} \in \mathbb{R}^Z$ is the hidden state at time $t$ and frequency $f$ output by the encoder. The function $\text{gate}(\boldsymbol{h})$ is a neural network consisting of a fully-connected layer, ReLU activation, another fully-connected layer, and a sigmoid function. $\tilde{\boldsymbol{h}}_{t,f} \in \mathbb{R}^Z$ is the resulting vector that sums $\boldsymbol{h}_{t,f}$ and $\boldsymbol{h}_{\text{sv}}$ based on $\boldsymbol{r}_{t,f}$, and this is passed to the decoder.

\subsection{APC Fine-Tuning}\label{subsec:ft}
Using the data described above, the AMT model is fine-tuned for the APC task.

For the six types of data output by the hFT-Transformer, cross-entropy (binary cross-entropy for onsets and frames) is measured for each element, and the loss is calculated accordingly.

\begin{align}\label{eq:loss-matrix}
L^{\text{matrix}}
  &= \text{CrossEntropy}(y, \hat y) \notag \\
  &= \frac{1}{512\times 88} \sum_{t,p,v} y_{t,p,v} \log \hat y_{t,p,v}
\end{align}

We calculate the loss for each hierarchy by summing the losses for these three types of data, and finally, the two losses are combined with a ratio of $\beta$ to obtain the final loss.

\begin{align}\label{eq:loss}
L_h &= \frac{1}{3}(L^{\text{onset}}_h + L^{\text{frames}}_h + L^{\text{velocity}}_h)\\
L &= \beta L_1 + (1 -\beta) L_2
\end{align}

In practice, the loss is not calculated for all elements in the matrix; instead, it is focused on particularly important elements. Calculating the loss for all elements can hinder effective learning. We consider the positions where the ground truth labels are non-zero, and their surrounding elements, to be the important ones. Specifically, we extract the positions $(t, p)$ where the ground truth labels are non-zero, as well as their neighboring pitches $(t, p-1)$ and $(t, p+1)$. The loss is then calculated using these elements and a randomly selected subset of remaining elements, which are sampled with a probability $\theta^{\text{matrix}}$. The model parameters are updated based on this calculation.

\section{Experiment}\label{sec:ex}
\subsection{Dataset}\label{subsec:dataset}
For fine-tuning, we created a dataset of paired original audio tracks and piano-covered MIDI files. The dataset creation followed the method of pop2piano, where original audio tracks and piano cover audio pairs were sourced from YouTube. These pairs were synchronized using synctoolbox \cite{synctoolbox}, and MIDI files were extracted using the AMT model. Based on Uta-Net’s \cite{utanet} rankings, we selected popular songs and obtained their piano covers from YouTube. We then automatically and manually removed unsuitable covers (e.g., covers in a different key, poor audio quality), resulting in a final dataset of 332 songs and 1,267 covers.

\subsection{Training}\label{subsec:train}
We trained the model on 80\% (265 songs) of the prepared piano covers. The Adam \cite{adam} optimizer was used, with a learning rate set to 1e-4. For parameters not explicitly stated, we adopted the default values provided by PyTorch \cite{pytorch}. As for the loss-related parameters, we used $\beta = 0.75, \theta^{\text{onset}} = 0.07, \theta^{\text{frame}} = 0.2, \theta^{\text{velocity}} = 0.01$. The training was conducted using two NVIDIA A100 GPUs, with a batch size of 8, over 5 epochs

Training completed in approximately 3 hours. On the test data, which consisted of the remaining 20\% of the songs not used during training, the average F1 score for onset, frame, and velocity was 0.31 (with onset and frame considered positive where the threshold exceeded 0.5, and velocity where the argmax was non-zero).

\subsection{Reproducibility of Original Song}\label{subsec:ex-rep}
We evaluated how accurately the piano covers generated by the APC model could reproduce the original songs. As an evaluation metric, we used the similarity between the original track and the cover, calculated by a Cover Song Identification (CSI) model. The CSI task involves determining whether a song is a cover, and the CSI model is considered capable of accurately calculating the similarity between the original audio and the cover.

First, we generated piano covers for a total of 67 original tracks included in the test data. The style vector was randomly chosen from covers that were not used in the training process. After removing notes shorter than 0.08 seconds from the MIDI, the MIDI was converted to audio using a synthesizer. We then calculated the similarity between the original track and the generated audio. The synthesizer used was FluidSynth with the FluidR3\_GM soundfont. For similarity, we adopted the $\mathcal Q_{\text{max}}$ metric \cite{qmax}, and the Python implementation of ChromaCoverID \cite{chromacoverid} was used for the actual computation. The $\mathcal Q_{\text{max}}$ value is based on chroma features, with smaller values indicating higher similarity between the two audio tracks.

Table \ref{tab:qmax} summarizes the $\mathcal Q_{\text{max}}$ values for the APC model trained with AMT-APC and other APC models. For each APC model, piano covers were generated for all songs in the test dataset, and the average $\mathcal Q_{\text{max}}$ value with the original track was calculated. The "Human" row represents piano covers from the test data that were converted to MIDI and then re-synthesized using the same soundfont, with $\mathcal Q_{\text{max}}$ calculated in the same way. As shown in Table \ref{tab:qmax}, the APC model trained with AMT-APC achieved the smallest $\mathcal Q_{\text{max}}$ value, indicating that it better reproduced the characteristics of the original songs compared to other models.

\setlength{\tabcolsep}{10pt}
\begin{table}[tb]
  \caption{Average $\mathcal Q_{\text{max}}$ for each APC model.}
  \begin{center}
  \begin{tabular}{c c}
    \hline
    Model & Average $\mathcal Q_{\text{max}}$ \\ \hline
    Human & 0.038 \\
    by AMT-APC (Proposed) & 0.035 \\
    pop2piano \cite{pop2piano} & 0.090 \\
    PiCoGen2 \cite{picogen2} & 0.054 \\ \hline
  \end{tabular}
  \label{tab:qmax}
  \end{center}
\end{table}

\subsection{Influence of Style Vector}\label{subsec:ex-sv}
\begin{figure}[tb]
  \centerline{\includegraphics[width=\linewidth]{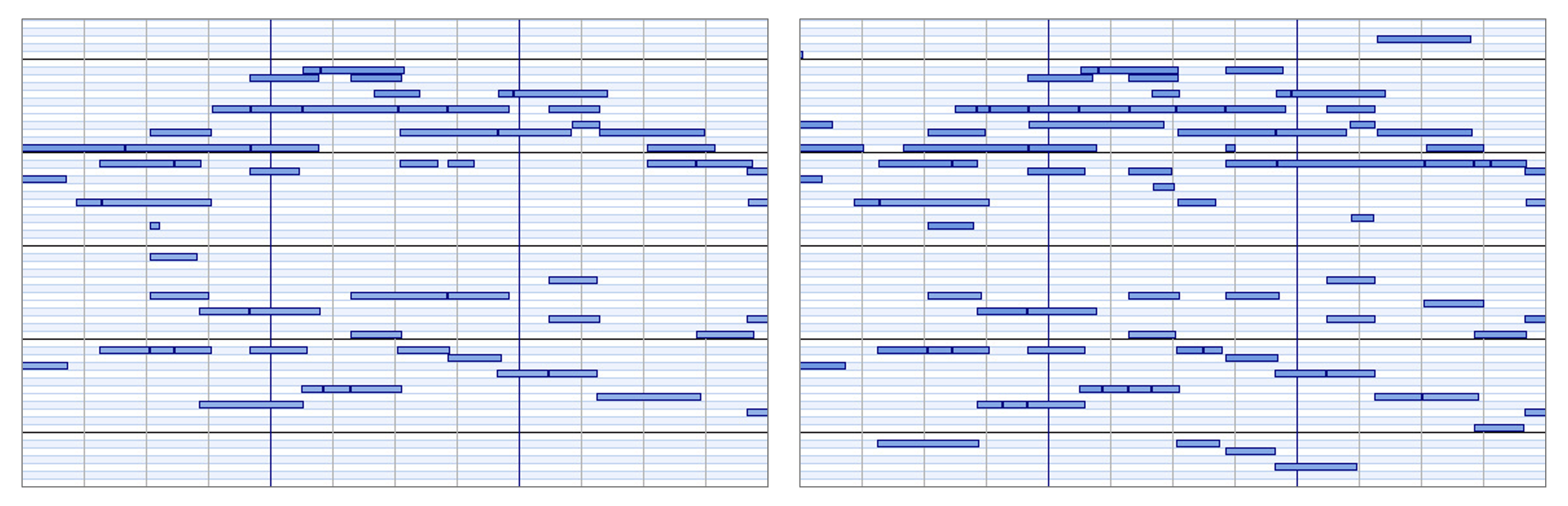}}
  \caption{Differences in piano covers generated by the style vector. Left: Calm style. Right: Intense style.}
  \label{fig:ressv}
\end{figure}

We examined the influence of the style vector on the generated MIDI. As an example, Figure \ref{fig:ressv} shows two covers generated with different styles. We selected several calm and intense performances based on the amount of notes and averaged the style vectors obtained from those performances before feeding them into the model. The left cover represents the calm style, while the right cover represents the intense style. It is evident that the number of notes and the hand movements differ between the two.

\subsection{Ablation Study}\label{subsec:ex-ab}
We conducted an ablation study to compare the performance when omitting AMT pre-training and the style vector from the training process. All other training parameters were the same as those described in section \ref{subsec:train} Additionally, we tested a raw AMT model without APC fine-tuning. The results are shown in Table \ref{tab:abl}. From Table \ref{tab:abl}, it can be seen that performance decreases when AMT fine-tuning or the style vector is excluded from the training process. Furthermore, the raw AMT model still exhibited a certain level of performance when applied directly to APC. The F1 score of the raw AMT model was 0.12, which is comparable to the score of a model trained for 2-3 epochs without AMT pre-training.

\begin{table}[tb]
  \caption{Average loss and F1 score for each model.}
  \begin{center}
  \begin{tabular}{c c c}
    \hline
    Model & loss & Average F1 \\ \hline
    Full & 0.36 & 0.31 \\
    w/o AMT pre-training & 0.41 & 0.27 \\
    w/o APC fine-tuning & 0.92 & 0.12 \\
    w/o Style vector & 0.43 & 0.30 \\ \hline
  \end{tabular}
  \label{tab:abl}
  \end{center}
\end{table}

\section{Discussion}\label{discussion}
The experiments in section \ref{subsec:ex-ab} demonstrated that AMT pre-training accelerates the learning process for APC. Additionally, it was found that a well-trained AMT model already possesses a modest ability to perform APC. This suggests that AMT and APC are highly similar tasks, and developing a robust AMT model is key to acquiring an effective APC model.

When training the APC model without incorporating the style vector, a slight decrease in performance was observed. This indicates that the style vector helps reduce confusion in the model. As shown in Figure \ref{fig:ressv}, the difference in output due to the variation in the style vector demonstrates that the model effectively expresses the features provided by the style vector.

One of the challenges of the proposed method is that it struggles to maintain strong consistency throughout the entire piece. While this issue is partially addressed by the style vector, the style vector we proposed only contains broad information, such as velocity and note density, and lacks detailed features like accompaniment patterns and ornamentations. As a result, consistency in these elements cannot be maintained.

\section{Conclusion}\label{conclusion}
In this study, we proposed AMT-APC, a learning algorithm for APC models that leverages pre-developed AMT models. We demonstrated that the APC model trained using this method can reproduce original tracks more accurately than any existing model. This research suggests that AMT and APC are highly similar tasks, and maximizing the use of existing AMT research results brings us closer to developing superior APC models. Further development of this research is anticipated, such as discovering or constructing AMT models better suited for APC applications, to achieve even more advanced APC models.

\section*{Acknowledgment}
In the writing of this paper, GPT-4o was used for translation into English.

\bibliographystyle{IEEEtran}
\bibliography{references}

@article{transformer,
  title={Attention is all you need},
  author={Vaswani, A},
  journal={Advances in Neural Information Processing Systems},
  year={2017}
}

@article{GPT1,
  title={Improving language understanding by generative pre-training},
  author={Radford, A},
  year={2018}
}

@article{BERT,
  title={Bert: Pre-training of deep bidirectional transformers for language understanding},
  author={Devlin, Jacob},
  journal={arXiv preprint arXiv:1810.04805},
  year={2018}
}

@article{GAN,
  title={Generative adversarial nets},
  author={Goodfellow, Ian and Pouget-Abadie, Jean and Mirza, Mehdi and Xu, Bing and Warde-Farley, David and Ozair, Sherjil and Courville, Aaron and Bengio, Yoshua},
  journal={Advances in neural information processing systems},
  volume={27},
  year={2014}
}

@inproceedings{Rombach2022,
  title={High-resolution image synthesis with latent diffusion models},
  author={Rombach, Robin and Blattmann, Andreas and Lorenz, Dominik and Esser, Patrick and Ommer, Bj{\"o}rn},
  booktitle={Proceedings of the IEEE/CVF conference on computer vision and pattern recognition},
  pages={10684--10695},
  year={2022}
}

@article{jukebox,
  title={Jukebox: A generative model for music},
  author={Dhariwal, Prafulla and Jun, Heewoo and Payne, Christine and Kim, Jong Wook and Radford, Alec and Sutskever, Ilya},
  journal={arXiv preprint arXiv:2005.00341},
  year={2020}
}

@article{music-transformer,
  title={Music transformer},
  author={Huang, Cheng-Zhi Anna and Vaswani, Ashish and Uszkoreit, Jakob and Shazeer, Noam and Simon, Ian and Hawthorne, Curtis and Dai, Andrew M and Hoffman, Matthew D and Dinculescu, Monica and Eck, Douglas},
  journal={arXiv preprint arXiv:1809.04281},
  year={2018}
}

@inproceedings{REMI,
  title={Pop music transformer: Beat-based modeling and generation of expressive pop piano compositions},
  author={Huang, Yu-Siang and Yang, Yi-Hsuan},
  booktitle={Proceedings of the 28th ACM international conference on multimedia},
  pages={1180--1188},
  year={2020}
}

@inproceedings{multitrack-music-transformer,
  title={Multitrack music transformer},
  author={Dong, Hao-Wen and Chen, Ke and Dubnov, Shlomo and McAuley, Julian and Berg-Kirkpatrick, Taylor},
  booktitle={ICASSP 2023-2023 IEEE International Conference on Acoustics, Speech and Signal Processing (ICASSP)},
  pages={1--5},
  year={2023},
  organization={IEEE}
}

@article{onsets-frames,
  title={Onsets and frames: Dual-objective piano transcription},
  author={Hawthorne, Curtis and Elsen, Erich and Song, Jialin and Roberts, Adam and Simon, Ian and Raffel, Colin and Engel, Jesse and Oore, Sageev and Eck, Douglas},
  journal={arXiv preprint arXiv:1710.11153},
  year={2017}
}

@article{MT3,
  title={MT3: Multi-task multitrack music transcription},
  author={Gardner, Josh and Simon, Ian and Manilow, Ethan and Hawthorne, Curtis and Engel, Jesse},
  journal={arXiv preprint arXiv:2111.03017},
  year={2021}
}

@article{hawthorne2021,
  title={Sequence-to-sequence piano transcription with transformers},
  author={Hawthorne, Curtis and Simon, Ian and Swavely, Rigel and Manilow, Ethan and Engel, Jesse},
  journal={arXiv preprint arXiv:2107.09142},
  year={2021}
}

@inproceedings{hft-transformer,
    author={Keisuke Toyama and Taketo Akama and Yukara Ikemiya and Yuhta Takida and Wei-Hsiang Liao and Yuki Mitsufuji},
    title={Automatic Piano Transcription with Hierarchical Frequency-Time Transformer},
    booktitle={Proceedings of the 24th International Society for Music Information Retrieval Conference},
    year={2023}
}

@inproceedings{pop2piano,
  title={Pop2Piano: Pop audio-based piano cover generation},
  author={Choi, Jongho and Lee, Kyogu},
  booktitle={ICASSP 2023-2023 IEEE International Conference on Acoustics, Speech and Signal Processing (ICASSP)},
  pages={1--5},
  year={2023},
  organization={IEEE}
}

@article{T5,
  title={Exploring the limits of transfer learning with a unified text-to-text transformer},
  author={Raffel, Colin and Shazeer, Noam and Roberts, Adam and Lee, Katherine and Narang, Sharan and Matena, Michael and Zhou, Yanqi and Li, Wei and Liu, Peter J},
  journal={Journal of machine learning research},
  volume={21},
  number={140},
  pages={1--67},
  year={2020}
}

@inproceedings{picogen1,
  title={PiCoGen: Generate Piano Covers with a Two-stage Approach},
  author={Tan, Chih-Pin and Guan, Shuen-Huei and Yang, Yi-Hsuan},
  booktitle={Proceedings of the 2024 International Conference on Multimedia Retrieval},
  pages={1180--1184},
  year={2024}
}

@inproceedings{picogen2,
    author = {Tan, Chih-Pin and Ai, Hsin and Chang, Yi-Hsin and Guan, Shuen-Huei and Yang, Yi-Hsuan},
    title = {PiCoGen2: Piano cover generation with transfer learning approach and weakly aligned data},
    year = 2024,
    month = nov,
    booktitle = {Proceedings of the 25th International Society for Music Information Retrieval Conference (ISMIR)},
    address = {San Francisco, CA, United States},
}

@article{kong2021,
  title={High-resolution piano transcription with pedals by regressing onset and offset times},
  author={Kong, Qiuqiang and Li, Bochen and Song, Xuchen and Wan, Yuan and Wang, Yuxuan},
  journal={IEEE/ACM Transactions on Audio, Speech, and Language Processing},
  volume={29},
  pages={3707--3717},
  year={2021},
  publisher={IEEE}
}

@article{kim2019,
  title={Adversarial learning for improved onsets and frames music transcription},
  author={Kim, Jong Wook and Bello, Juan Pablo},
  journal={arXiv preprint arXiv:1906.08512},
  year={2019}
}

@article{synctoolbox,
  title={Sync Toolbox: A Python package for efficient, robust, and accurate music synchronization},
  author={M{\"u}ller, Meinard and {\"O}zer, Yigitcan and Krause, Michael and Pr{\"a}tzlich, Thomas and Driedger, Jonathan},
  journal={Journal of Open Source Software},
  volume={6},
  number={64},
  pages={3434},
  year={2021}
}

@article{sheetsage,
  title={Melody transcription via generative pre-training},
  author={Donahue, Chris and Thickstun, John and Liang, Percy},
  journal={arXiv preprint arXiv:2212.01884},
  year={2022}
}

@inproceedings{MAESTRO,
  title={Enabling Factorized Piano Music Modeling and Generation with the {MAESTRO} Dataset},
  author={Curtis Hawthorne and Andriy Stasyuk and Adam Roberts and Ian Simon and Cheng-Zhi Anna Huang and Sander Dieleman and Erich Elsen and Jesse Engel and Douglas Eck},
  booktitle={International Conference on Learning Representations},
  year={2019},
  url={https://openreview.net/forum?id=r1lYRjC9F7},
}

@article{musemorphose,
  title={MuseMorphose: Full-song and fine-grained piano music style transfer with one transformer VAE},
  author={Wu, Shih-Lun and Yang, Yi-Hsuan},
  journal={IEEE/ACM Transactions on Audio, Speech, and Language Processing},
  volume={31},
  pages={1953--1967},
  year={2023},
  publisher={IEEE}
}

@misc{utanet,
  author={{Uta-Net}},
  title={Artist Ranking},
  howpublished={\url{https://www.uta-net.com/ranking/artist/}},
  note={Accessed: 2024-08-10, in Japanese}
}

@article{adam,
  title={Adam: A method for stochastic optimization},
  author={Kingma, Diederik P},
  journal={arXiv preprint arXiv:1412.6980},
  year={2014}
}

@article{pytorch,
  title={Pytorch: An imperative style, high-performance deep learning library},
  author={Paszke, Adam and Gross, Sam and Massa, Francisco and Lerer, Adam and Bradbury, James and Chanan, Gregory and Killeen, Trevor and Lin, Zeming and Gimelshein, Natalia and Antiga, Luca and others},
  journal={Advances in neural information processing systems},
  volume={32},
  year={2019}
}

@article{qmax,
  title={Cross recurrence quantification for cover song identification},
  author={Serra, Joan and Serra, Xavier and Andrzejak, Ralph G},
  journal={New Journal of Physics},
  volume={11},
  number={9},
  pages={093017},
  year={2009},
  publisher={IOP Publishing}
}

@misc{chromacoverid,
  author={Albin Correya and Kyle McDonald},
  title={ChromaCoverID},
  howpublished={\url{https://github.com/albincorreya/ChromaCoverId}},
  note={Accessed: 2024-08-21}
}

\end{document}